\documentclass[twocolumn,a4paper]{aastex701}
\usepackage[T1]{fontenc}
\usepackage{amsmath,bm}
\usepackage{graphicx}
\usepackage[svgnames]{xcolor}
\usepackage{tikz}  
\usepackage{csquotes}
\usepackage{appendix}
\usepackage{cleveref}
\crefname{figure}{Figure}{Figures}
\crefname{section}{Section}{Sections}
\crefname{equation}{Equation}{Equations}
\newcommand*\dd{\mathop{}\!\mathrm{d}}

\shorttitle{Lense--Thirring induced repeating blast trains in continuous gravitational waves}
\shortauthors{K. R. Doney, S. Jana, O. A. Hannuksela, P. Martens}

\begin{document}

\title{Lense--Thirring induced repeating blast trains in continuous gravitational waves}

\author{Kailib Ryan Doney}
\email{1155229952@link.cuhk.edu.hk}
\affiliation{Department of Physics, The Chinese University of Hong Kong, Shatin, New Territories, Hong Kong}

\author{Souvik Jana}
\email{souvikjana@cuhk.edu.hk}
\affiliation{Department of Physics, The Chinese University of Hong Kong, Shatin, New Territories, Hong Kong}

\author{Otto A. Hannuksela}
\email{oahannuksela@cuhk.edu.hk}
\affiliation{Department of Physics, The Chinese University of Hong Kong, Shatin, New Territories, Hong Kong}

\author{Paul Martens}
\email{paulmartens@cuhk.edu.hk}
\affiliation{Department of Physics, The Chinese University of Hong Kong, Shatin, New Territories, Hong Kong}

\begin{abstract}
    We characterise gravitational lensing induced by the Lense--Thirring effect in a continuous gravitational-wave source orbiting a spinning supermassive black hole.
    This scenario is very likely to generate lensing events, as long as the source's orbital inclination is not aligned with the black hole's spin axis and the Lense--Thirring timescale is shorter than the observational window.
    The signal appears as short-duration ($T_\text{blast}$) gravitational-wave \enquote{blasts} that repeat after every orbit ($T_\text{orb}$) within a limited time window ($T_\text{train}$).
    The blasts' signal strength depends on the system's geometry and not on the black hole mass.
    The signal will go silent and renew twice in a full Lense--Thirring precession period.
\end{abstract}


\section{Introduction}
\label{sec:intro}


The Lense--Thirring effect, also known as inertial frame dragging, makes the plane of an orbit precess about the spin axis of the central body \citep{LenseThirring1918, Mashhoon1984, Ciufolini2004, Ciufolini:2023czv}.
Near \objectname{Sgr~A$^*$}, the Lense--Thirring precession of stars and pulsars on orbits closer than the known S-stars has been proposed as a test of the Kerr metric \citep{Kerr1963}, through stellar astrometry or radio pulsar timing \citep{Will2008, Liu2012, Psaltis2016}.
Because of the Lense--Thirring effect, if an object follows an orbital motion around a massive body whose spin axis is misaligned with the orbit by an angle $\theta_\mathrm{i}$, the orbital plane will precess, while keeping its inclination $\theta_\mathrm{i}$ fixed (see Figure \ref{fig:Lens-Thirring_diagram}).
Searches for gravitational lensing of gravitational waves have so far focused on compact binary coalescences \citep{Hannuksela2019, LVC2021Lensing, LVK2024Lensing, GWTC4Lensing2026} observed by Advanced LIGO \citep{aLIGO2015}, Advanced Virgo \citep{aVirgo2015} and KAGRA \citep{KAGRA2021}.
The Galactic Centre is expected to host many neutron stars, and directed searches have looked for their continuous gravitational waves in Advanced LIGO and Advanced Virgo data \citep{Piccinni2020, LVK2022GC}.
A neutron star behind \objectname{Sgr~A$^*$} would be strongly lensed into two interfering images, a signal that third-generation detectors may observe \citep{Basak2023} and characterise even beyond the Einstein radius \citep{Savastano2024}, while lenses within the Galaxy can also magnify continuous waves from Galactic sources \citep{Suyamprakasam2025}. Repeated gravitational-wave lensing of sources on short orbits around a supermassive black hole has been studied in hierarchical triples \citep{DOrazio:2019fbq,Yu:2021dqx}, and Lense--Thirring precession has been shown to modulate that repeated lensing for continuous-wave sources in active galactic nucleus disks \citep{Li:2025xuh}.
Building on that work, we characterise the resulting blast trains for a continuous-wave source around \objectname{Sgr~A$^*$}: closed timescale and mean-gain formulae, and the finding that the time-averaged magnification barely raises the integrated signal-to-noise ratio.
\citet{Li2026} showed that Lense--Thirring precession modulates the repeated lensing of continuous-wave sources orbiting supermassive black holes in AGN discs, where the precession period is much longer than the observation and slowly changes successive lensing events.
Here we consider the opposite regime around \objectname{Sgr~A$^*$}, where the precession period is shorter than the observation.
We show that if the orbiting body emits continuous gravitational waves, the Lense--Thirring effect will orient the system in such a way that the gravitational waves are lensed by the central massive body in a periodic repeating pattern, causing a `train' of amplified gravitational-wave `blasts' that switches on and off twice per Lense--Thirring period.

\section{Lense--Thirring precession near \texorpdfstring{Sgr~A$^*$}{Sgr A*}}
\label{sec:lt}

\begin{figure*}[t]
    \centering
    \includegraphics{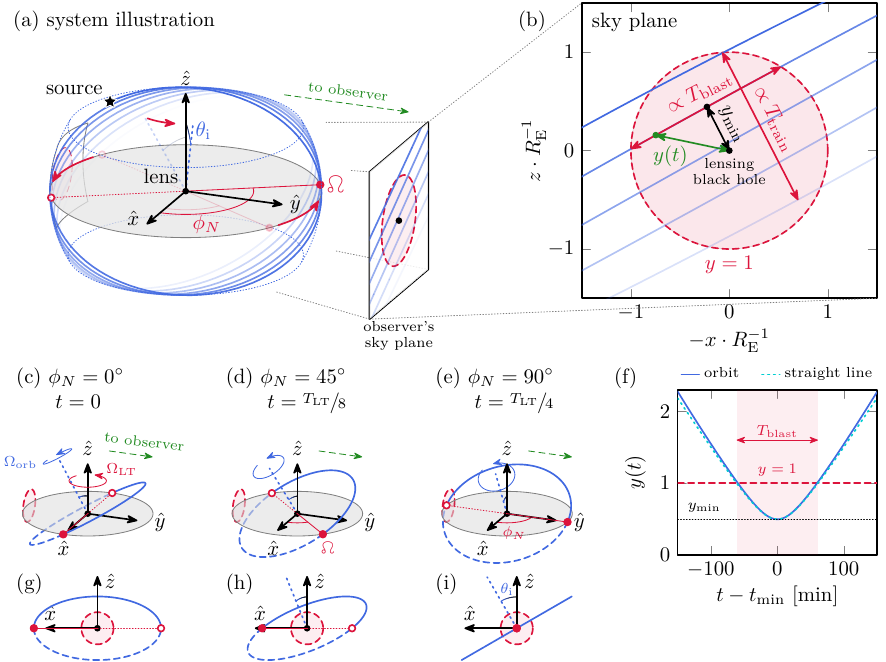}
    \caption{
        Illustration of how the Lense--Thirring effect produces lensing.
        (a) The orbital plane of a continuous gravitational-wave source inclined at $\theta_\mathrm{i}=30^\circ$ precesses around a black hole's spin axis ($\hat{z}$-axis) due to the Lense--Thirring effect until it is (near) edge-on with respect to the observer, when it is lensed once per orbit.
        The longitude of the ascending node (filled red circle) grows as $\phi_N = \Omega_\text{LT}\, t$, and the red arc shows how far the node has moved since $t=0$.
        The filled and open red dots mark the ascending and descending nodes, respectively.
        (b) pictures the observer's sky plane, and also defines the dimensionless image position $y(t)$ used in the text.
        The dashed red circle ($y=1$) is the \enquote{Einstein radius}, which, approximately, marks where lensing occurs.
        (f) shows the time evolution of $y(t)$.
        (c), (d) and (e) give the orbital plane at given times, while the corresponding (g), (h) and (i) show what the observer would be seeing at these specific times.
        These are similar to (a) and (b).
        At $t = \frac{T_\text{LT}}{4}$ (e) and (i), the orbit is edge-on and passes through the Einstein ring, producing a lensing blast once per orbit.
        These blasts repeat until the Lense--Thirring effect sweeps the orbit out of alignment on a timescale $T_\text{train}$ (b).
        Four timescales apply here: the Lense--Thirring timescale $T_\text{LT}$, the orbital timescale $T_\text{orb}$, the time for the source object to cross the Einstein radius once and produce a blast, $T_\text{blast}$, and the time for Lense--Thirring precession to sweep the orbit across the Einstein radius, $T_\text{train}$.
        See also \cref*{fig:timescales}.
}
    \label{fig:Lens-Thirring_diagram}
\end{figure*}

In the weak-field limit, a point-like mass follows a quasi-circular orbit of radius $R$ around a black hole of mass $M$ with an angular velocity $\Omega_\text{orb}$.
Because of the Lense--Thirring effect, the line of nodes precesses around the central black hole's spin axis with dimensionless spin $a$ at a nodal frequency $\Omega_\text{LT}$.
In the weak-field limit, these angular frequencies can be expressed as \citep{LenseThirring1918, Bardeen1972, Will2008}
\begin{subequations}
    \begin{align}
        \Omega_\text{orb} & = \left(\frac{M}{R^3}\right)^{1/2}\,, \\
        \Omega_\text{LT}& = \frac{2aM^2}{R^3}\,,
    \end{align}
    \label{eq:timescales}
\end{subequations}
where we use geometrized units ($G = c = 1$).
Without spin, the orbital plane is fixed: the continuous gravitational wave source either never passes behind the black hole, or does so on every orbit in the same way.
With spin, the line of nodes precesses about the spin axis, and its longitude grows as $\phi_N = \Omega_\text{LT}\, t$ (\cref{fig:Lens-Thirring_diagram}).
At any instant, the source still follows a quasi-circular path in a fixed plane at
the current $\phi_N$; we call this plane the \emph{instantaneous orbit}.
Once every orbit, the source is behind the black hole, producing strong
lensing (a \enquote{blast}).

\section{Will lensing occur?}
Whether Lense--Thirring induced lensing occurs depends on the timescale versus
observation time.
For a source orbiting
$\tilde r = 100$ gravitational radii ($\tilde r=R/M$)
around a rapidly spinning black hole ($a=0.9$) with the mass of
\objectname{Sgr~A$^*$} ($M = 4.3 \times 10^{6}\,M_\odot$)
\citep{Gravity2022}, the orbital and Lense--Thirring periods are
\begin{align}
    T_\text{orb} &
        \approx 1.5\,\mathrm{day}
        \left(\frac{\tilde r}{100}\right)^{\frac{3}{2}}
        \left(\frac{M}{4.3\times10^{6}\,M_\odot}\right) \, ,
    \label{eq:Torb_scaling} \\
    T_\text{LT} &
        \approx 2.3\,\mathrm{yr}
        \left(\frac{\tilde r}{100}\right)^{3}
        \left(\frac{0.9}{a}\right)
    \left(\frac{M}{4.3\times10^{6}\,M_\odot}\right) \, ,
    \label{eq:TLT_scaling}
\end{align}
so the orbital plane precesses once in about $560$ orbits.
To see the
precession within an observing time $T_\text{obs}$, we need $T_\text{LT} <
T_\text{obs}$, which limits the orbital radius in gravitational radii to
\begin{equation}
    \tilde r \lessapprox 130 \left( \frac{a}{0.9} \right)^{\frac{1}{3}}
    \left( \frac{T_\text{obs}}{5\,\mathrm{yr}} \right)^{\frac{1}{3}}
    \left( \frac{M}{4.3\times10^{6}\,M_\odot} \right)^{-\frac{1}{3}} \, .
    \label{eq:xmax}
\end{equation}
For $T_\text{obs} = 5\,\mathrm{yr}$, our example covers about two Lense--Thirring
periods, so the instantaneous orbit is perfectly edge-on to the line of sight about
four times.
We call each such instant the \emph{edge-on time}.
This does not mean that the source object is then perfectly behind the black hole:
a dead-centre passage requires the edge-on time to coincide with a far-side node
crossing, so the Lense--Thirring and orbital timescales must align.
Under these conditions, the optical depth is
\begin{equation}
    \tau(\theta_\mathrm{i})  \approx \sin\theta_\mathrm{i} \,.
\end{equation}
Indeed, in this parameter space, the Lense--Thirring effect practically guarantees that strong lensing occurs whenever the observer lies within $\theta_\mathrm{i}$ of the black hole equatorial plane, the band of directions swept by the orbital plane (Appendix~\ref{app:observer}).\footnote{This does not account for detector selection, only the geometric probability of lensing.}
Precession similarly raises the lensing probability for sources in AGN discs \citep{Li2026}.

\begin{figure*}[t]
    \centering
    \includegraphics[width=\textwidth]{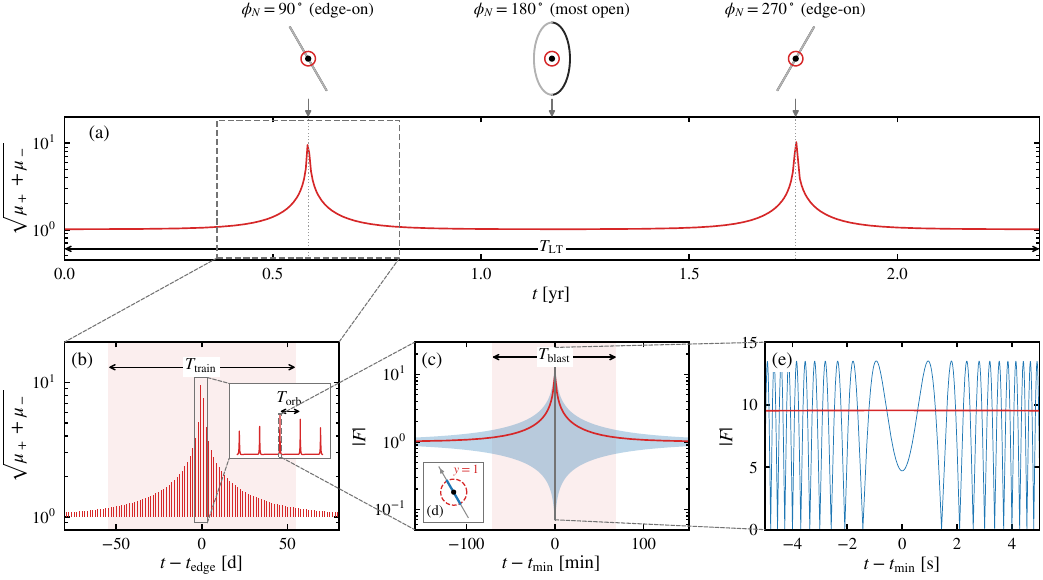}
    \caption{
    Time scales of the blasts for the system of \cref*{fig:Lens-Thirring_diagram}, at $f_\text{GW} = 1\,\mathrm{kHz}$. Top: the orbit as seen by the observer at $\phi_N = 90^\circ$, $180^\circ$, and $270^\circ$; dark: behind the black hole; red circle: largest Einstein radius.
    (a) The root-mean-square amplification $\sqrt{\mu_+ + \mu_-}$ at the closest approach of each orbit, over one Lense--Thirring period, $T_\text{LT} = 2.3\,\mathrm{yr}$; the two peaks are the two blast trains.
    (b) The first train: one line per orbit; shaded: $y_\text{min} < 1$, which lasts $T_\text{train} = 111\,\mathrm{day}$ ($72$ blasts). Inset: five blasts around the best orbit, $T_\text{orb} = 1.5\,\mathrm{day}$ apart.
    (c) The best blast ($y_\text{min} = 0.011$): the root-mean-square (red) and the range of $|F|$ over the fringes, $\sqrt{\mu_+} \pm \sqrt{\mu_-}$ (blue); shaded: $y < 1$, $T_\text{blast} = 140\,\mathrm{min}$.
    (d) Its path in the lens plane, in units of the local Einstein radius.
    (e) $|F|(t)$ within $5\,\mathrm{s}$ of closest approach, showing the interference of the two images (blue) and the root-mean-square (red). Grey boxes and dashed lines show each zoom; the boxes in (b) and (c) are as wide as the next panel and so appear as lines.}
    \label{fig:timescales}
\end{figure*}


\section{A gravitational-wave blast from an Einstein crossing}
\label{sec:single_blast}

When the orbiting source object enters the Einstein radius of the lensing black hole,
the waves are lensed strongly, producing a short \enquote{blast} of gravitational waves on
a timescale $T_\text{blast}$.
Given the large distances between the objects, we treat the lensing black hole
as a point lens in the thin-lens approximation \mbox{\citep{Schneider1992}}, and the
orbiting body producing continuous gravitational waves as a point source.
The Einstein radius is \citep{Einstein1936, Schneider1992}
\begin{equation}
    R_\text{E} = \sqrt{\frac{4 M D_L D_{LS}}{D_S}} \approx \sqrt{4 M D_{LS}} \,,
    \label{eq:einstein}
\end{equation}
where $D_L$, $D_S$ and $D_{LS}$ are the angular diameter distances from the
observer to the lens and the source, and from the lens to the
source, with $D_{LS} \ll D_L \approx 8.3\,\mathrm{kpc}$ for
\objectname{Sgr~A$^*$} \citep{Gravity2022}.
Approximating the Einstein radius as a constant near the lensing region, such that the
distance from the lens to the source object is approximately the orbital radius, $D_{LS} \approx R$,
\begin{equation}
    R_\text{E} \approx
        0.85\,\mathrm{AU} \,
        \left(\frac{\tilde r}{100}\right)^{\frac{1}{2}} \,
        \left(\frac{M}{4.3\times10^{6}\,M_\odot}\right) \, .
    \label{eq:RE_scaling}
\end{equation}
When the binary enters this radius ($y \equiv b/R_\text{E} < 1$), the
gravitational-wave amplitude
is greatly amplified (by a factor of approximately $\sqrt{\mu_+ + \mu_-}$, where $\mu_\pm$
are the magnifications of the two images
\cite{Schneider1992}),
producing a short-duration blast of gravitational waves of duration
\begin{equation}
    T_\text{blast} = 2\,t_\text{E}\sqrt{1 - y_\text{min}^2} \, ,
    \label{eq:tblast}
\end{equation}
where $t_\text{E}=R_\text{E}/v\approx 2 \tilde r M$ is the Einstein crossing
time \citep{Paczynski1986}.
This $T_\text{blast}$ is the Einstein-radius crossing duration of one orbital pass (denoted $T_\text{Ein}$ in \citealt{Li:2025xuh}); here we give the closed chord formula.

\footnote{\Cref{eq:tblast} assumes a straight-line
    approximation with no orbital curvature, which holds within 0.5\,\%
    accuracy (\cref{fig:timescales}). 
}
As a rough scaling, for \objectname{Sgr~A$^*$},
\begin{align}
    T_\text{blast} &\approx 140\,\mathrm{min} \left (\frac{\tilde r}{100}\right)\left(\frac{M}{4.3\times10^{6}\,M_\odot}\right)\sqrt{1 - y_\text{min}^2} \, .
    \label{eq:tblast_scaling}
\end{align}
The source spends a duration
\begin{equation}
    \begin{split}
        T_\text{peak} &= 2\sqrt{3}\,t_\text{E}\,y_\text{min} \\
        &\approx 245\,\text{min}\left(\frac{\tilde r}{100}\right)\left(\frac{M}{4.3\times10^{6}\,M_\odot}\right) y_\text{min} \, ,
    \end{split}
    \label{eq:Tpeak}
\end{equation}
inside the high-magnification region covering around 50\% of the peak, $y < 2 y_\text{min}$.\footnote{From the straight-line chord $y(t)^2 = y_\text{min}^2 + ((t-t_0)/t_\text{E})^2$.}
These relations estimate the duration of the fast gravitational-wave blasts
near the caustic; we use them below for the strongest blast in a train.

\section{From a single blast to a train of blasts}
\label{sec:trains}

After the orbit crosses the Einstein radius,
it resumes its motion until the next orbit and
it crosses the Einstein radius again, producing another blast.
This train of blast signals repeats over several cycles.
A similar orbit-to-orbit modulation appears for sources in AGN discs \citep{Li2026}, although there an observation covers only a small part of a precession period.
Each orbit crosses the Einstein ring a little
further along (\cref{fig:timescales}), 
closer to a perfectly edge-on scenario, though unfortunately it is
unlikely that the source ever passes dead-centre behind the black hole.
The orbital node advances by a Lense--Thirring advance
$\Delta \phi_N=\Omega_\text{LT} T_\text{orb} = {4\pi a}/{\tilde r^{\frac{3}{2}}}$ in the azimuthal direction per orbit.
Converting this into the angle between the line of
sight and the point where the orbiting gravitational-wave source passes closest
to the dead-centre behind the black hole,\footnote{Near edge-on,
    $\sin\beta=|\hat{L}\cdot\hat{n}|$ with $\hat{L}$ the unit orbital angular
    momentum and $\hat{n}$ the line of sight, so a nodal advance $\Delta\phi_N$
    changes $\beta$ by $\sin\theta_\mathrm{i}\,\Delta\phi_N$.
}
we have the closest encounter at each far-side node crossing near the edge-on time,
\begin{align}
    y_\text{min}(n) &\approx 
    \frac{2\pi a \sin\theta_\mathrm{i}}{\tilde r}\,\left|n - \frac{t_\text{edge}}{T_\text{orb}}\right| \, ,
    \label{eq:ymin_n}
\end{align}
where the nodal integer $n=t_\text{node}/T_\text{orb}$, choosing the time origin so that
node crossings fall on integer multiples of the orbital time, while $t_\text{edge}$ is the
time at which the \enquote{instantaneous orbit} is perfectly edge-on to the line of
sight.\footnote{The source does not necessarily pass dead-center behind the black hole at this time.}
Successive crossings happen once every orbit, so the closest encounter $y_\text{min}$ shifts by
$\Delta y_\text{min}=2\pi a \sin\theta_\mathrm{i}/\tilde r$ from one blast to the
next.\footnote{The closest blast in a train therefore has $y_\text{min} \lesssim \pi a\sin\theta_\mathrm{i}/\tilde r$.}
A single train of blasts contains the orbits inside the Einstein radius
($y_\text{min} < 1$),
so the total number of blast cycles and the duration of one train is given
by\footnote{That is, $\left|n - t_\text{edge}/T_\text{orb}\right| < \frac{\tilde r}{2\pi a\sin\theta_\mathrm{i}}$ on each side of the edge-on time.}
\begin{align}
    N_\text{cycles} &
    = \frac{\tilde r}{\pi a \sin\theta_\mathrm{i}} \, ,
    \nonumber\\
    T_\text{train} &= N_\text{cycles}\,T_\text{orb} = \frac{2\tilde r^{\frac{5}{2}} M}{a\sin\theta_\mathrm{i}} \, .
    \label{eq:Ttrain}
\end{align}
As a rough scaling relation,
\begin{align}
    N_\text{cycles} &\approx 71\left(\frac{\tilde r}{100}\right)\left(\frac{0.9}{a}\right)\left(\frac{0.5}{\sin\theta_\mathrm{i}}\right) \, ,
    \label{eq:Ncycles_scaling}\\
    T_\text{train} &\approx 110\,\mathrm{day}\left(\frac{\tilde r}{100}\right)^{\frac{5}{2}}\left(\frac{0.9}{a}\right)
    \left(\frac{0.5}{\sin\theta_\mathrm{i}}\right)
    \nonumber\\
    &\quad\times\left(\frac{M}{4.3\times10^{6}\,M_\odot}\right) \, .
    \label{eq:Ttrain_scaling}
\end{align}
The system can therefore be described as follows --
On a single Lense--Thirring timescale, the orbiting source object will produce
two trains of gravitational-wave blasts lasting $T_\text{train}$,
with each train containing $N_\text{cycles}$ blasts separated
by orbital time $T_\text{orb}$ and each lasting $T_\text{blast}$ (\cref{fig:timescales}).
The closest the source object gets to the black hole during a transit in dimensionless
source position
$y_\text{min} \sim \pi a\sin\theta_\mathrm{i}/\tilde r$,
which can be used to estimate the maximal magnification produced by the
strongest blast in the train.

\section{Peak of the loudest blast}
\label{sec:fastest_blast}

At every orbit during a train of blasts, the source object sweeps past the dead-centre line of sight behind the black hole with a known nearest source position $y_\text{min}$ (\cref{eq:ymin_n}).
The loudest blast comes from the far-side node crossing closest to the edge-on time, offset from it by $\Delta_\text{edge} \equiv \min_n \left|n - t_\text{edge}/T_\text{orb}\right| \leq \frac{1}{2}$ orbits, so its nearest source position is
\begin{equation}
    y_\text{nearest}
    \approx 0.014
        \left(\frac{a}{0.9}\right)
        \left(\frac{\sin\theta_\mathrm{i}}{0.5}\right)
        \left(\frac{100}{\tilde r}\right)
        \left(2\Delta_\text{edge}\right)
        \, ,
\end{equation}
which gives a peak duration of the loudest blast (enclosing 50\% of the peak magnification) of
\begin{equation}
    \begin{split}
        T_\text{peak} \approx 3.5\,\mathrm{min} &\left(\frac{M}{4.3\times10^{6}\,M_\odot}\right)\left(\frac{a}{0.9}\right) \\
        &\times\left(\frac{\sin\theta_\mathrm{i}}{0.5}\right)\left(2\Delta_\text{edge}\right) \, .
    \end{split}
    \label{eq:Tpeak_best}
\end{equation}
The very peak of the gravitational-wave blast depends sensitively on the orbit in the Lense--Thirring cycle.
However, the duration of the very peak of the magnification / closest encounter is very brief,
lasting only an order minute for typical parameters considered here.
The timescale is, interestingly, independent of the orbital radius $\tilde r$.

\section{Lense--Thirring induced blast amplification}
\label{sec:amplification}

In geometric optics limit,\footnote{Geometric optics holds for $wy \gg 1$.}
the black hole splits the source object signal into two gravitational-wave \enquote{images}
that in the continuous-wave limit interfere, such that the original gravitational-wave polarisations
$h_{+,\times}$ are convolved by an amplification factor \citep{Takahashi2003}
\begin{equation}
    F = \sqrt{\mu_+}\,e^{i\phi_+} + \sqrt{\mu_-}\,e^{i\phi_-}, \quad
    \mu_\pm = \frac{1}{2}\left(\frac{y^2+2}{y\sqrt{y^2+4}} \pm 1\right) \, ,
    \label{eq:F}
\end{equation}
where the two images are time-shifted and Doppler-modulated with a frequency difference of $\pm 2f_\text{GW}/\tilde{r}$,\footnote{$\pm 20\,\mathrm{Hz}$ at $1\,\mathrm{kHz}$ for \objectname{Sgr~A$^*$} at $\tilde r = 100$.}
such that the two images beat at a frequency of $4f_\text{GW}/\tilde{r}$.
During a blast, the two images interfere, so
the magnification due to the amplification factor oscillates between
the two different magnifications ($\sqrt{\mu_+} - \sqrt{\mu_-}$ and $\sqrt{\mu_+} + \sqrt{\mu_-}$) (\cref{fig:timescales}).
One can estimate roughly the average magnification at a given
source position $y$ by taking the mean-square of the
amplification factor (assuming high-magnification limit $y\ll 1$):
\begin{equation}
    \langle|F|^2\rangle = \langle \mu \rangle \approx \frac{1}{y}\,. 
\end{equation}
The oscillation is slow at closest approach,
where the time delay between the
images is stationary, and reaches about $40\,\mathrm{Hz}$ away from it
(\cref{fig:timescales}).  
Therefore, twice every Lense-Thirring cycle, 
the image is split into two potentially visible images with 
slight frequency modulation and rapidly magnified. 

\section{Time-averaged magnification over a train of blasts}
\label{sec:successive_blasts}

After a single blast, the source continues to orbit, producing 
a second wave of gravitational-wave blasts that can be 
characterised by the same amplification factor $F$ (\cref{eq:F}), 
where the entire train of blasts can be described using 
time-averaging. 
Successive orbits cross the lens plane on parallel lines spaced by
$\Delta y_\text{min}$.
At the closest approach ($y=y_\text{min}$), the mean-square of the amplification factor
reaches its maximum
\begin{equation}
    \langle|F|^2\rangle_\text{max}
    \,\approx 70 \left(\frac{\tilde r}{100}\right)\left(\frac{0.9}{a}\right)\left(\frac{0.5}{\sin\theta_\mathrm{i}}\right) \, .
    \label{eq:Fmax}
\end{equation}
At typical frequencies ($f_\text{GW} = 1\,\mathrm{kHz}$),
and closest encounter of $y_\text{min} \sim 10^{-2}$,
the blast is still far below the diffraction scale,
so diffraction effects are not relevant (Appendix~\ref{app:wave}).
For an observer in the black hole
equatorial plane, the time spent per unit area of the lens plane is
approximately uniform.
For $T_\text{obs} \geq T_\text{LT}$,
the fraction of time spent at magnification $\mu = \mu_+ + \mu_- > \mu_0$
(from the inversion of \cref{eq:F})
\begin{equation}
    f(\mu_0)
    = \frac{4}{\pi\tilde{r}\sin\theta_\mathrm{i}}
      \left(\frac{\mu_0}{\sqrt{\mu_0^2 - 1}} - 1\right)
    \,.
    \label{eq:duty}
\end{equation}
For example, in the typical scenario that we have been considering,\footnote{$\tilde r=100$, $a=0.9$, $\theta_\mathrm{i}=30^\circ$}
the fraction of time the source object spends inside the Einstein radius
is around $1.3\,\%$.
Integrating the exact magnification over the lens plane gives the time-averaged power gain,
\begin{equation}
    \langle|F|^2\rangle - 1 = \frac{4}{\pi\tilde{r}\sin\theta_\mathrm{i}} \, , \label{eq:gain}
\end{equation}
which is 2.5\,\% in our example.
Interestingly, the fractional time spent in a blast (\Cref{eq:duty}) and the
time-averaged power gain (\cref{eq:gain}) do not depend on the black hole $M$, spin $a$ or the
gravitational-wave frequency $f_\text{GW}$.
Therefore, if the Lense--Thirring timescale is shorter than the observation time,
the average magnification boost depends only on the orbital radius and inclination.

\section{Does lensing boost detectability?}
\label{sec:detectability}

A rotating neutron star on an inclined orbit close to a spinning
black hole produces fast gravitational-wave burst repeaters, with some
of those bursts reaching magnifications up to $100$ or even more (depending
on the orbital and Lense--Thirring timescales), but the effect does
not help in detecting these sources due to the short duration of each burst.
Specifically, suppose that the source object is a neutron star with equatorial ellipticity
$\epsilon$ and moment of inertia $I$, with $f_\text{GW}$ equal to twice its
spin frequency, at distance $d$.
In that case, its strain amplitude
\begin{equation}
    h_0 = \frac{4\pi^2 G I \epsilon f_\text{GW}^2}{c^4 d} \, .
    \label{eq:h0}
\end{equation}
To a reasonable degree of accuracy, the amplitude $h_0$ and the inclinations
$\theta_\mathrm{i}$ are constant over the observation time $T_\text{obs}$.
As for other lensed continuous-wave sources near the Galactic Centre \citep{Basak2023}, we consider a third-generation detector, the Einstein Telescope \citep{Branchesi2023}.
Averaged over the star's spin orientation and the Earth's rotation, in the absence of lensing, the semi-coherent signal-to-noise ratio of the gravitational-wave emission recorded in the Einstein Telescope is
\begin{equation}
    \rho_0^2 = \frac{9}{25}\frac{h_0^2 T_\text{coh} N}{S_n(f_\text{GW})} \, , \label{eq:rho0}
\end{equation}
where $S_n$ is the noise power spectral density of one of the three
interferometers and $T_\text{coh}$ is the coherent integration time of each segment, with $N$ segments in total.
Such semi-coherent searches analyse each segment coherently and add the segment powers without tracking the phase between segments, trading sensitivity for computational cost \citep{Tenorio2021, Riles2023, Wette2023}.
Once we apply lensing, the semi-coherent signal magnitude
changes such that
\begin{equation}
    \rho_\text{L}^2 =
    \rho_0^2 \sqrt{\frac{\int|F|^2\,\dd t}{T_\text{obs}}} =
    \rho_0^2 \sqrt{\langle|F|^2\rangle} \, ,
\end{equation}
where the timing model is assumed to include the Doppler modulation
between the two images.
Therefore, a rough scaling for the lensed signal-to-noise ratio
\begin{equation}
    \frac{\rho_\text{L}}{\rho_0}
        \approx 1 + 0.006\left(\frac{100}{\tilde r}\right)^{1/4} \left(\frac{0.5}{\sin\theta_\mathrm{i}}\right)^{1/4}
        \, .
        \label{eq:rhoL}
\end{equation}
That is, despite the high-magnification bursts,
the lensing barely increases the overall signal-to-noise
ratio at all.\footnote{If the search is coherent, then the signal-to-noise ratio scales as
$\frac{\rho_\text{L}}{\rho_0} \approx 1 + 0.0125
\left(\frac{100}{\tilde r}\right)^{1/2}
\left(\frac{0.5}{\sin\theta_\mathrm{i}}\right)^{1/2}$, though the conclusion remains the same.}
The reason for this counter-intuitive result is that the
lensing at a high magnification
only occurs for a brief peak period (\cref{eq:Tpeak_best}) of the total
observation time, while the non-lensed signal is present for years.
This integrated-SNR conclusion complements work that instead asks whether the lensing or spin imprint is distinguishable in the waveform \citep{Li:2025xuh}.
In fact, the detection of these blasts may be more difficult
than the detection of the actual continuous-wave signal due to
the nature of the gravitational-wave analysis.
The most likely scenario is that the continuous-wave signal
is detected first, and then the blasts are identified as lensed.
These lensing effects do not make a typical continuous gravitational-wave
source more detectable.

\section{Example system: \texorpdfstring{Sgr~A$^*$}{Sgr A*} with a neutron star}

We take the system shown in \cref{fig:Lens-Thirring_diagram} at
distance $d=8.3\,\mathrm{kpc}$, with gravitational-wave frequency
$f_\text{GW}=1\,\mathrm{kHz}$, ellipticity $\epsilon=10^{-7}$,
moment of inertia $I=1\times10^{38}\,\mathrm{kg\,m^2}$ and
observation time $T_\text{obs}=5\,\mathrm{yr}$.
This ellipticity is at the low end of the theoretical crustal maximum of about
$1\times10^{-7}$--$7\times10^{-6}$ \citep{Gittins2021,MoralesHorowitz2022}.
Over five years the observation covers about four trains and 300 blasts.
From \cref{eq:h0} to \cref{eq:rhoL}, the unlensed amplitude is
$h_0=1.3\times10^{-26}$, so $\rho_0\approx 170$ and $\rho_\text{L}\approx 21$,
both well above a detection threshold of $\rho=8$.
As in Section~\ref{sec:detectability}, this assumes a timing model that
follows the Doppler modulation over $T_\text{obs}$.

The orbital speed $v\approx 0.1$ produces a large Doppler shift:
$\pm 100\,\mathrm{Hz}$ over each orbit, with a drift of
$\dot{f}\approx f_\text{GW}\,(\tilde{r}^2M)^{-1}\approx4.7\times10^{-3}\,\mathrm{Hz\,\mathrm{s}^{-1}}$
near closest approach behind the black hole.
Without a timing model, the coherent integration time is then limited to about
$15\,\mathrm{s}$, since the signal must stay within one frequency bin of width $1/T_\text{coh}$ \citep{Riles2023}.
Directed searches of the Galactic Centre \citep{Piccinni2020, LVK2022GC} cover frequency derivatives many orders of magnitude smaller than this, so a source on such an orbit would need a dedicated search that follows the orbit.
At that coherence time, a semi-coherent search of the blasts alone stays
below 1 standard deviation (Appendix~\ref{app:stats}).
With a timing model that follows a full blast of about $2.3\,\mathrm{h}$,
the same search reaches about 19 standard deviations, against a threshold of 10.
No single blast is detectable on its own: even the loudest stays below
$\rho = 8$.
A semi-coherent search over all of $T_{\text{obs}}$ at the same
$2.3\,\mathrm{h}$ coherence time gives about 75 standard deviations,
because the blasts occupy only about 1\,\% of the data.
As in Section~\ref{sec:detectability}, the continuous-wave signal would
therefore be found first, and the blasts would identify it as lensed
afterwards.
Sgr~A$^*$ is within reach for this signal if the timing model can follow
the Doppler modulation across a blast.

\subsection{Expected sources}

Gravitational-wave inspiral from $\tilde{r}=100$ takes about $4\,\mathrm{Myr}\,(\tilde{r}/100)^4(1.4\,M_\odot/m)$ \citep{Peters1964}.

With a neutron-star inspiral rate of order $1\times10^{-8}\,\mathrm{yr^{-1}}$ per galaxy \citep{HopmanAlexander2006,AmaroSeoanePreto2011}, we expect about $0.02$ to $0.2$ neutron stars this close to Sgr~A$^*$, before requiring a fast spin or a nearly circular orbit.

\section{Conclusions}
\label{sec:conclusions}

If such a Lense--Thirring system exists around Sgr~A$^*$, its signal may be observable.
We focused on an observer in the black hole equatorial plane, for which any inclined orbit with $T_\text{LT}<T_\text{obs}$ becomes edge-on twice per Lense--Thirring period, giving two blast trains spaced by $T_\text{LT}/2$.
Other observer orientations change the signal rhythm, giving unevenly spaced or merged trains (Appendix~\ref{app:observer}).
We have focused on the geometric optics case.
Blasts that approach the diffraction limit are possible, but at $1\,\mathrm{kHz}$ they occur in only about $10^{-4}$ of trains (Appendix~\ref{app:wave}).

Relative to Lense--Thirring modulation of repeated continuous-wave lensing in active galactic nucleus disks \citep{Li:2025xuh}, our emphasis is the Galactic-centre blast-train structure: analytic $T_\text{blast}$, $T_\text{train}$, and $y_\text{min}$ relations, a time-averaged power gain that depends only on orbital radius and inclination (\cref{eq:gain}), and the result that this gain barely raises the integrated signal-to-noise ratio.
The source would therefore be found first as an ordinary continuous-wave signal, with the blasts identifying it as lensed.

We also treated the source as a neutron star seen only in gravitational waves.
If it were also a radio pulsar, pulse timing could supply the orbital solution \citep{Liu2012, Psaltis2016}, making the fully coherent analysis realistic. However, scattering towards Sgr~A$^*$ makes such timing difficult.
The same alignments would also lens the radio pulses, giving an independent check of the blast times.
A detection would also probe the black hole itself, because the orbital period, the spacing of the trains and their length fix $\tilde{r}$, $a$ and $\theta_\mathrm{i}$, giving a gravitational-wave measurement of the spin of Sgr~A$^*$.

\begin{acknowledgments}
We thank Aditya Sharma for reviewing the manuscript.
This material is based upon work supported by NSF's LIGO Laboratory which is a major facility fully funded by the National Science Foundation.
The authors are grateful for computational resources provided by the LIGO Laboratory and supported by National Science Foundation Grants PHY-0757058 and PHY-0823459.
This paper carries LIGO document number LIGO-P2600604.
\end{acknowledgments}

\bibliography{references}
\bibliographystyle{aasjournalv7}

\begin{appendices}
\section{Wave optics and the diffraction limit} \label{app:wave}

\subsection{The diffraction limit}

The geometric-optics amplification in \cref{eq:F} diverges as $y\to0$, but the signal's wave nature keeps it finite.
The relevant parameter is the dimensionless frequency $w=8\pi M(1+z_{\text{L}})f_{\text{GW}}\approx8\pi Mf_{\text{GW}}$.
For $y\ll1$ and $w\gg1$, the amplification of a point lens is \citep{Takahashi2003}
\begin{equation}
    |F|^2 \approx \pi w\,J_0^2(wy) \, , \label{eq:wave_F}
\end{equation}
where $J_0$ is a Bessel function.
At $y=0$ this gives the diffraction limit,
\begin{equation}
    |F|_{\text{max}} = \sqrt{\pi w} \approx 1300\left(\frac{M}{4.3\times10^{6}\,M_\odot}\right)^{1/2}\left(\frac{f_{\text{GW}}}{1\,\mathrm{kHz}}\right)^{1/2} \, . \label{eq:diff_limit}
\end{equation}
For $wy\gg1$, \cref{eq:wave_F} oscillates between 0 and $\frac{2}{y}$ with a mean of $\frac{1}{y}$, which reproduces the fringes and root-mean-square amplification of geometric optics at small $y$.
Wave effects therefore matter only for $y\lesssim w^{-1}$, which is $1.9\times10^{-6}$ at $1\,\mathrm{kHz}$, far below the $y_{\text{min}}\sim10^{-2}$ of the blasts in the main text.

\subsection{Wave-optics blasts}

\begin{figure}[t]
    \centering
    \includegraphics{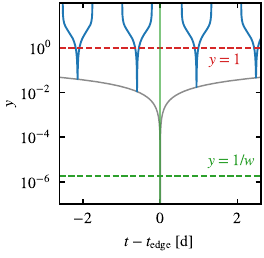}
    \caption{The source object near the first edge-on time of \cref*{fig:timescales}, at $f_\text{GW} = 1\,\mathrm{kHz}$ ($w = 5.3\times10^5$). Blue: $y(t)$ along the orbit, with dips $T_\text{orb} = 1.5\,\mathrm{day}$ apart.
    Grey: The $y_\text{min}$ of a passage at time $t$.
    Dashed: $y = 1$ and $y = w^{-1}$.
    Green: the diffraction window ($T_\text{diffrac} = 18\,\mathrm{s}$, too narrow to resolve).
    The deepest dip, $0.6\,\mathrm{day}$ before the edge-on time, has $y_\text{min} = 0.011 \gg w^{-1} = 1.9\times10^{-6}$.}
    \label{fig:wave}
\end{figure}

Since $y_{\text{min}}$ grows linearly near the edge-on time (\cref{eq:ymin_n}), the condition $y_{\text{min}}\lesssim w^{-1}$ holds only in a short diffraction window,
\begin{equation}
    T_{\text{diffrac}}
        = \frac{T_{\text{train}}}{w}
        = \frac{\tilde{r}^{5/2}}{4\pi f_{\text{GW}}\,a\sin\theta_\mathrm{i}}
        \approx
            18\,\mathrm{s}
            \left(\frac{\tilde{r}}{100}\right)^{\frac{5}{2}}
            \left(\frac{0.9}{a}\right)
            \left(\frac{0.5}{\sin\theta_\mathrm{i}}\right)
            \left(\frac{f_{\text{GW}}}{1\,\mathrm{kHz}}\right)^{-1} \, . \label{eq:Tdiffrac}
\end{equation}
As this is much shorter than one orbit, a wave-optics blast needs a special alignment.
The source must cross the node on the far side of the black hole within $\frac{T_{\text{diffrac}}}{2}$ of the edge-on time,
\begin{equation}
    |t_{\text{node}}-t_{\text{edge}}| < \frac{T_{\text{diffrac}}}{2} \, , \label{eq:tnode}
\end{equation}
with $t_{\text{edge}}$ and $t_{\text{node}}$ as in \cref{eq:ymin_n}.
In our system no passage falls inside the window (\cref{fig:wave}), as the closest one comes $0.6\,\mathrm{d}$ before $t_{\text{edge}}$, with $y_{\text{min}}=0.011\gg w^{-1}$.
The number of orbits in the window is
\begin{equation}
    N^{\text{diffrac}}_{\text{cycles}}
        = \frac{T_{\text{diffrac}}}{T_{\text{orb}}}
        = \frac{\tilde{r}}{\pi w\,a\sin\theta_\mathrm{i}}
        \approx
            1.3\times10^{-4}
            \left(\frac{\tilde{r}}{100}\right)
            \left(\frac{0.9}{a}\right)
            \left(\frac{0.5}{\sin\theta_\mathrm{i}}\right)
            \left(\frac{M}{4.3\times10^{6}\,M_\odot}\right)^{-1}
            \left(\frac{f_{\text{GW}}}{1\,\mathrm{kHz}}\right)^{-1} \, . \label{eq:Ndiffrac}
\end{equation}
A wave-optics blast therefore occurs in about $10^{-4}$ of trains, or with a probability of about $6\times10^{-4}$ in $5\,\mathrm{yr}$ for this system.
Every train reaches the wave-optics regime only if $N^{\text{diffrac}}_{\text{cycles}}\gtrsim1$, that is, $w\lesssim N_{\text{cycles}}$.
For Sgr~A$^*$ at $\tilde{r}=100$ this requires $f_{\text{GW}}\lesssim 0.13\,\mathrm{Hz}$, far below the Einstein Telescope band.

A wave-optics blast would also be brief.
The source crosses $y<w^{-1}$ in $\frac{2t_\text{E}}{w}\approx16\,\mathrm{ms}$ at $1\,\mathrm{kHz}$, about 16 gravitational-wave cycles.
Using \cref{eq:duty}, it spends a fraction $\frac{2}{\pi\tilde{r}w^2\sin\theta_\mathrm{i}}\approx4.5\times10^{-14}$ of the time in this regime, so wave optics does not affect the results of Section~\ref{sec:detectability}.

\subsection{Effect of black hole spin}

\cref{eq:wave_F} assumes a point caustic.
The spin of a Kerr black hole unfolds this point into a small astroid \citep{SerenoDeLuca2008}, with a half-width in units of $R_{\text{E}}$ of about
\begin{equation}
    y_{\text{c}}
        \approx \frac{15\pi}{512}\frac{a^2}{\tilde{r}^{3/2}}
        \approx
            7.5\times10^{-5}
            \left(\frac{a}{0.9}\right)^2
            \left(\frac{\tilde{r}}{100}\right)^{-\frac{3}{2}} \, ,
\end{equation}
for an observer in the black hole equatorial plane.
This is 40 times larger than $w^{-1}$ at $1\,\mathrm{kHz}$, so the astroid is resolved at this frequency.
A ring-diffraction estimate then gives a peak of about $|F|\approx400$, so \cref{eq:diff_limit} is an upper limit.
The point-lens value applies only below about $25\,\mathrm{Hz}$.

\section{Observer orientation} \label{app:observer}

The main text assumes the observer lies in the black hole equatorial plane.
Here we let the observer lie at a latitude $\delta$ above that plane.
The angle $\beta$ between the line of sight and the orbital plane then follows
\begin{equation}
    \sin\beta = |\cos\delta\sin\theta_\mathrm{i}\cos\phi_N - \sin\delta\cos\theta_\mathrm{i}| \, , \label{eq:beta_general}
\end{equation}
which reduces to $\sin\theta_\mathrm{i}|\cos\phi_N|$ for $\delta=0$.
A far-side passage at angle $\beta$ has $b_\text{min} = R\sin\beta$ and $D_{LS} = R\cos\beta$, so \cref{eq:einstein} gives $y_{\text{min}} = \frac{\sqrt{\tilde r}}{2}\,\frac{\sin\beta}{\sqrt{\cos\beta}}$.

\subsection{Train pattern}

For $|\delta|\leq\theta_\mathrm{i}$, the orbit is edge-on at $\phi_N=\pm\phi_*$, with
\begin{equation}
    \cos\phi_* = \frac{\tan\delta}{\tan\theta_\mathrm{i}} \, .
\end{equation}
Successive trains are therefore separated alternately by $\frac{\phi_*}{\pi}\,T_{\text{LT}}$ and $\left(1-\frac{\phi_*}{\pi}\right)T_{\text{LT}}$ for $\delta>0$, which gives the signal a short-long rhythm instead of the even spacing of $T_{\text{LT}}/2$.
The angle $\beta$ also changes more slowly near each edge-on time, so each train lengthens,
\begin{equation}
    T_{\text{train}}(\delta)
        \approx
        T_{\text{train}}(0)
        \left(1-\frac{\sin^2\delta}{\sin^2\theta_\mathrm{i}}\right)^{-\frac{1}{2}} \, ,
\end{equation}
and the number of blasts grows by the same factor.
This holds to within about 10\,\% for trains that remain separate.

Across the short gap, $\beta$ reaches only $\theta_\mathrm{i}-|\delta|$.
If this is below about $2\tilde{r}^{-1/2}$, $y_{\text{min}}$ stays below 1 throughout the gap and the two trains merge into one long train per Lense--Thirring period.
For $|\delta|>\theta_\mathrm{i}$, the orbit is never edge-on.
The closest passage has $\beta=|\delta|-\theta_\mathrm{i}$, giving a single train per period with a floor on $y_{\text{min}}$, and the blasts vanish once $|\delta|-\theta_\mathrm{i}\gtrsim2\tilde{r}^{-1/2}$.

For the system of \cref{fig:Lens-Thirring_diagram}, $2\tilde{r}^{-1/2}$ corresponds to $11.5^\circ$.
At $\delta=10^\circ$, trains of 78 blasts are separated alternately by $0.94\,\mathrm{yr}$ and $1.40\,\mathrm{yr}$.
The trains merge above $\delta\approx18.5^\circ$, so at $\delta=25^\circ$ there is one train of about 210 blasts per period, containing two edge-on times $0.47\,\mathrm{yr}$ apart.
At $\delta=35^\circ$, one train of about 130 blasts has $y_{\text{min}}\geq0.44$, which limits the root-mean-square amplification to about 1.6.
Above $\delta\approx41.5^\circ$ there are no blasts.
The ratio $\frac{\rho_{\text{L}}}{\rho_0}$ rises from 0.12 at $\delta=0$ to 0.16 at $\delta=28^\circ$ because the trains last longer, then falls to 0.09 at $\delta=35^\circ$.
The equatorial observer of the main text is therefore a conservative choice for $|\delta|<\theta_\mathrm{i}$.

\subsection{Likelihood of alignment}

For a static orbit, reaching $y<y_0$ requires the line of sight to lie within $\beta_0\approx2y_0\tilde{r}^{-1/2}$ of the orbital plane, which holds for a fraction $\sin\beta_0$ of observers.
This is only 0.3\,\% for $y_0=0.014$ at $\tilde{r}=100$.
Precession sweeps the orbital plane through every direction with $|\delta|\leq\theta_\mathrm{i}$, a fraction $\sin\theta_\mathrm{i}$ of the sky, so each train reaches the same alignment for half of all observers at $\theta_\mathrm{i}=30^\circ$.
Averaging over isotropic orbits gives $\frac{\pi}{4}$, or 79\,\%.
The aligned case is therefore not fine-tuned.

\section{Detection statistics} \label{app:stats}

\subsection{Coherent signal-to-noise ratio}

For one L-shaped interferometer, averaging over sky position and source orientation gives $\rho^2=(4/25)h_0^2T/S_n$.
The triangular design has three interferometers with $60^\circ$ opening angles, so the response scales by $3\sin^2 60^\circ$, giving the factor $9/25$ in \cref{eq:rho0}.
For a source at polar angle $\theta$ from the detector normal, the triangle response is proportional to $1+6\cos^2\theta+\cos^4\theta$.
Averaged over a sidereal day towards Sgr~A$^*$, this gives $\rho$ within 4\,\% of the sky-averaged value at both candidate sites, so \cref{eq:rho0} holds.
Within a day, $\rho$ varies by a factor of 0.56 to 1.55, which matters for a single blast but averages out when many segments are combined.

\subsection{Semi-coherent search}

A semi-coherent search divides the data into $N$ segments of length $T_{\text{coh}}$, analyses each coherently, and sums the power without tracking the phase between segments \citep{Riles2023, Tenorio2021, Wette2023}.
Each segment contributes a power $\rho_i^2$, and in noise the summed statistic has a variance of $8N$, so the significance is
\begin{equation}
    S = \frac{\sum_i\rho_i^2}{\sqrt{8N}} \, . \label{eq:stack}
\end{equation}
For a search of the blasts alone, $N=f(3/\sqrt{5})\,T_{\text{obs}}/T_{\text{coh}}$ and $\rho_i^2=(9/25)(h_0^2/S_n)\,\bar{\mu}\,T_{\text{coh}}$, where $\bar{\mu}=\sqrt{5}\approx2.2$ is the mean of $\mu_++\mu_-$ over $y<1$.
Compared with a search over all of $T_{\text{obs}}$ at the same coherence time, this gives
\begin{equation}
    \frac{S_{\text{blast}}}{S_{\text{all}}} = f(3/\sqrt{5})^{1/2}\,\bar{\mu} \approx 0.25 \, ,
\end{equation}
for the system of \cref{fig:Lens-Thirring_diagram}.
The summed power of the example signal is about 830 for any $T_{\text{coh}}$, so $N$ sets the significance.
With $T_{\text{coh}}=2.3\,\mathrm{h}$, $N\approx240$ and $S\approx19$.
With $T_{\text{coh}}=15\,\mathrm{s}$, $N\approx1.4\times10^5$ and $S\approx0.8$.
A search over all of $T_{\text{obs}}$ with $T_{\text{coh}}=2.3\,\mathrm{h}$ gives $S\approx75$.

The $S=10$ threshold is indicative and does not include trial factors from searching over the orbit.
The search also assumes a template that follows both images within each segment.

\section{Approximations} \label{app:approx}

\paragraph{Weak-field lensing.} At the Einstein radius the deflection angle is $2\tilde{r}^{-1/2}=0.2\,\mathrm{rad}$ for $\tilde{r}=100$, so the thin-lens, weak-field treatment is only approximate.
The second-order deflection increases $R_{\text{E}}$ by a fraction $(15\pi/64)\tilde{r}^{-1/2}$, or about 7\,\%, and $T_{\text{blast}}$ by the same amount.
This shifts the timescales but does not change the train structure.

\paragraph{Black hole spin in the lens.} In the weak-deflection limit, the spin displaces the lens centre by $aM$ \citep{SerenoDeLuca2008}, which is $a/(2\tilde{r}^{1/2})\approx0.045$ in units of $R_{\text{E}}$.
This moves the deepest blast of a train by about one orbit.
We discuss the astroid effect in Appendix~\ref{app:wave}.

\paragraph{Quadrupole precession.} The spin-induced quadrupole changes $\Omega_{\text{LT}}$ by a fraction $\frac{3a\cos\theta_\mathrm{i}}{4\tilde{r}^{1/2}}\approx6\%$.
We neglect it here, but a measurement of $a$ from $T_{\text{LT}}$ would need to include it.

\paragraph{Orbital shape and decay.} We take the orbit to be circular.
An eccentric orbit would undergo apsidal precession of $6\pi/\tilde{r}$ per orbit, or $10.8^\circ$ with a period of $51\,\mathrm{d}$, modulating the orbital radius at closest approach behind the black hole and thus the blast duration across a train.
Gravitational-wave emission shrinks $\tilde{r}$ by only a fraction of about $3\times10^{-7}$ in $5\,\mathrm{yr}$.

\paragraph{Spin of the star.} Geodetic precession rotates the star's spin by $3\pi/\tilde{r}$ per orbit, or $5.4^\circ$ with a period of $103\,\mathrm{d}$, which is close to $T_{\text{train}}$.
Unless the spin is aligned with the orbit, this modulates the signal amplitude across a train.
Gravitational-wave spin-down at $\epsilon=10^{-7}$ and $1\,\mathrm{kHz}$ shifts $f_{\text{GW}}$ by only about $3\,\mathrm{mHz}$ in $5\,\mathrm{yr}$, so we take $h_0$ as constant.

\end{appendices}

\end{document}